# Probabilistic method to determine electron correlation energy

T.R.S. Prasanna

Department of Metallurgical Engineering and Materials Science

Indian Institute of Technology, Bombay

Mumbai – 400076 India

A new method to determine electron correlation energy is described. This method is based on a better representation of the potential due to interacting electrons that is obtained by specifying both the average and standard deviation. The standard deviation is determined from a probabilistic interpretation of the Coulomb interaction between electrons. This leads to a better representation of orbital energies as $\varepsilon_i \pm \Delta\varepsilon_i$, where $\varepsilon_i$ is the Hartree-Fock orbital energy and $\Delta\varepsilon_i$, the spread, is an indicator of the magnitude of correlation energy. This new representation of the potential when combined with an empirical constant leads naturally to a new method to determine electron correlation energy. Correlation energy is determined within the independent electron approximation without any contribution from higher energy unoccupied states. A consistent physical interpretation - an electron occupies a given position when other electrons are farther than on average – can be made. It is a general technique that can be used to determine correlation energy in any system of particles with inter-particle interaction V(**r₁, r₂**) and can be considered to be universal first step beyond mean-field theory.

The problem of determining electron correlation energy is of great importance in quantum chemistry and solid state physics. The origin of this problem can be traced to the Hartree-Fock approximation. It is an independent electron approximation in which the instantaneous electron-electron repulsion, $1/|\mathbf{r}_1 - \mathbf{r}_2|$, is replaced by an averaged electron-electron interaction in the Hamiltonian. Consequently, the Hartree-Fock ground state has a higher energy than the true ground state and the difference is defined as the correlation energy. This problem was recognized very early in the development of quantum mechanics as applied to molecules and solids. Subsequently, many techniques have been developed to address this problem. Configuration Interaction, Moller-Plesset Perturbation theory, Coupled Cluster method etc. have been developed to account for electron correlation in molecules. These methods are discussed in detail in Refs.1, 2. In these techniques, the Hartree-Fock ground state (single Slater determinant) is first determined, which then becomes a starting point to determine electron correlation energy. In general, it is necessary to include contributions from higher energy unoccupied states in these methods. An important consequence is that the independent electron approximation is no longer valid and the physically appealing picture of an electron represented by its wavefunction is lost. Additionally, these methods are computationally intensive as the number of Slater determinants increases rapidly and are impractical for solids.

Alternate approaches to address electron correlation within the single electron approximation have been made. These include studies based on Density Functional theory and on Green's functions (GW approximation and its extensions). Density Functional Theory is an alternate method to calculate electronic structure (3-5). Within this approach, the local density approximation (LDA) is the standard method to incorporate exchange effects. Many studies have been made go beyond the LDA to better characterize the exchange-correlation hole and incorporate the effects of electron correlations (6-11). The GW approximation and its extensions (12-15) represent another method to obtain better energies. Two broad themes can be distinguished in these studies. The first is to develop better functionals to describe electron correlations and the second is to develop schemes that lower the quasi-particle energies. In general, pair-correlation functions or Green's functions play an important role in these studies.

In this paper, a new method to determine electron correlation energy is described. It is based on the fact that the potential due to interacting electrons fluctuates at any position. Specifying both the average and standard deviation - as opposed to just the average as in the Hartree-Fock method - better represents this fluctuating potential. The standard deviation can be determined from a probabilistic interpretation of the Coulomb repulsion between electrons. Starting from the Hartree-Fock approximation, this method determines the correlation energy within the independent electron approximation without any contribution from higher energy unoccupied states. It results in lower orbital energies. A consistent physical interpretation - an electron occupies a given position when other electrons are farther than on average – can be made. It is a general technique that can be used to determine correlation energy in any system of particles with inter-particle interaction V($\mathbf{r_1}, \mathbf{r_2}$).

The Hartree-Fock (HF) approximation is discussed in detail in Ref.1. The orbital energy in the HF approximation is given by (1)

$$\varepsilon_i = \langle \psi_i | f_i | \psi_i \rangle = h_i + \sum_{j \neq i} J_{ij} - \sum_{j \neq i} K_{ij} \tag{1}$$

where $f_i$ is the Fock operator and all symbols have their usual meaning. The HF ground state energy is given by

$$E_{HF} = \sum_i \varepsilon_i - \frac{1}{2} \sum_i \sum_j (J_{ij} - K_{ij}) \tag{2}$$

where the second term compensates for double counting and $J_{ii} = K_{ii}$. The HF ground state has a higher energy than the true ground state partly because the Coulomb integral overestimates the repulsion energy between two electrons. The Coulomb integral between two electrons is given (in atomic units) by

$$J_{ij} = \iint |\psi_i(\mathbf{r_1})|^2 \frac{1}{|\mathbf{r_1} - \mathbf{r_2}|} |\psi_j(\mathbf{r_2})|^2 d\mathbf{r_1}\, d\mathbf{r_2} = \int |\psi_i(\mathbf{r_1})|^2 d\mathbf{r_1} \int \frac{|\psi_j(\mathbf{r_2})|^2}{|\mathbf{r_1} - \mathbf{r_2}|} d\mathbf{r_2} \tag{3}$$

This leads to the familiar interpretation that an electron in orbital $\psi_i$ (henceforth referred to as electron "$i$") experiences an average potential at $\mathbf{r_1}$ due to electron in $\psi_j$ (electron "$j$") given by

$$V_j^{av}(\mathbf{r_1}) = \int \frac{|\psi_j(\mathbf{r_2})|^2}{|\mathbf{r_1} - \mathbf{r_2}|} d\mathbf{r_2} \tag{4}$$

This is equivalent to the expression in classical electrostatics for the potential at any point due to a continuous charge distribution, in this case $(-e)|\psi_j(\mathbf{r_2})|^2$. In a classical charge distribution, the potential is constant because the charge distribution is constant with respect to time. In the quantum mechanical case, electrons are point particles and only occupy various positions with probability $|\psi_j(\mathbf{r_2})|^2$. Hence, it is readily seen that the potential at any position is not constant but fluctuates with its average given by Eq. (4). A fluctuating potential (or any other fluctuating quantity) is better described by specifying its standard deviation in addition to the average. This can be achieved if a probabilistic interpretation of Eq. (4) is made. Because $|\psi_j(\mathbf{r_2})|^2$ is a probability density function and not a charge density function, the average potential in Eq. (4) can also be interpreted as the expectation value of $1/|\mathbf{r_1} - \mathbf{r_2}|$ at $\mathbf{r_1}$. With this interpretation, it becomes possible to determine the variance at $\mathbf{r_1}$. The variance is given by

$$V_j^{\sigma^2}(\mathbf{r_1}) = \int \frac{|\psi_j(\mathbf{r_2})|^2}{|\mathbf{r_1} - \mathbf{r_2}|^2} d\mathbf{r_2} - \left(V_j^{av}(\mathbf{r_1})\right)^2 \tag{5}$$

Thus the variance can be determined if the expectation value of $1/|\mathbf{r_1} - \mathbf{r_2}|^2$ (first term of Eq. (5)) can be evaluated. As the probability density function is known, higher moments can also be determined if necessary. It is now possible to better characterize the potential at $\mathbf{r_1}$ due to an electron "$j$". It can be represented as

$$V_j(\mathbf{r_1}) = V_j^{av}(\mathbf{r_1}) \pm V_j^{\sigma}(\mathbf{r_1}) \tag{6}$$

where $V_j^{\sigma}(\mathbf{r_1})$ is the standard deviation and is given by square root of the variance, $V_j^{\sigma^2}(\mathbf{r_1})$, determined from Eq. (5). In an n-electron system, the average potential at $\mathbf{r_1}$ due to n-1 electrons can be represented (as is well known) by the sum of average potentials due to each electron. However, it is possible using the method described above to estimate the variance as well. The total variance can be represented by a sum of individual variances and its square root, the standard deviation, is given by

$$V_{\Sigma'i}^{\sigma}(\mathbf{r_1}) = \sqrt{V_{\Sigma'i}^{\sigma^2}(\mathbf{r_1})} = \sqrt{\sum_{j \neq i} V_j^{\sigma^2}(\mathbf{r_1})} \tag{7}$$

where Σ'i indicates that the quantity has been obtained from contributions of all electrons $j \neq i$. Therefore, the potential due to the other (n-1) electrons at $\mathbf{r_1}$ can be represented as

$$V_i^{tot}(\mathbf{r_1}) = \left(\sum_{j \neq i} V_j^{av}(\mathbf{r_1})\right) \pm V_{\Sigma'i}^{\sigma}(\mathbf{r_1}) \qquad (8)$$

This is a better representation of the potential due to n-1 electrons at $\mathbf{r_1}$ than the Hartree-Fock approximation, which is just the first term in Eq. (8). Specifying the average and standard deviation is the norm in describing any quantity that exhibits a spread in values. Eq. (8) appears to be the first time it has been done in the context of potential due to interacting electrons. As electrons occupy different positions, the distance to $\mathbf{r_1}$ and hence the potential at $\mathbf{r_1}$ fluctuates. Eq. (8) accounts for the fluctuation in the potential in a statistical manner.

Using Eq. (8) (instead of Eq. (4)) to calculate the potential energy of interacting electrons leads to a better representation of the orbital energy as $\varepsilon_i \pm \Delta \varepsilon_i$, where $\varepsilon_i$ is the orbital energy (same as the HF orbital energy obtained from Eq. (1)) and $\Delta \varepsilon_i$ is the spread. The method to determine $\Delta \varepsilon_i$ is described further below. The spread, $\Delta \varepsilon_i$, gives an estimate of the range of values about $\varepsilon_i$ that orbital energy can possess and is also an indicator of the magnitude of correlation energy. A large $\Delta \varepsilon_i$ follows from a large standard deviation of the potential and implies strong fluctuations about the average (HF) potential. This suggests a significant difference between the average (HF) potential and the true potential, indicating a large value of the correlation energy. Hence, extending the Hartree-Fock method to determine the spread of the orbital energy, $\Delta \varepsilon_i$, will provide an indication of the magnitude of correlation energy.

The true value of the orbital energy is a constant that does not exhibit any spread in the sense described above. This is because even though the potential at (any position) $\mathbf{r_1}$ fluctuates, the true value of the potential at $\mathbf{r_1}$ due to n-1 other electrons when electron '$i$' is present is a fixed quantity that is determined by the correlated motion of electrons. To determine this true potential, it is necessary to adopt theoretical techniques starting with the true many-body wavefunction, which is frequently unknown.

However, using Eq. (8) along with an empirical constant allows an effective potential to be estimated that is closer to the true value of the potential than the average (HF) potential. This naturally leads to a new method to determine electron correlation energy as described below. Electron "$i$" would prefer to occupy position $r_1$ when the potential is lower than on average as it would lower the (repulsive) energy. The effective potential at $r_1$ when electron "$i$" is present can be represented as

$$V_i^{eff}(\mathbf{r_1}) = \left(\sum_{j \neq i} V_j^{av}(\mathbf{r_1})\right) - c_i(\mathbf{r_1}) V_{\Sigma'i}^{\sigma}(\mathbf{r_1}) \tag{9}$$

The effective potential, given by Eq. (9), is closer to the true value of the potential at $r_1$ when electron "$i$" is present than the Hartree-Fock potential. It also implies that electron "$i$" occupies position $r_1$ when the other electrons are farther than on average. The coefficient $c_i(\mathbf{r_1})$ is a small number multiplying the standard deviation of the total potential due to other electrons and the representation is sufficiently general. The simplest assumption would be that of a constant value ($c$) for all electrons at all positions. The next assumption would be that of a different constant value for different electrons ($c_i$) but independent of position. Another possibility is to have one constant for electrons of same spin and another constant for electrons of opposite spin, as electrons of same spin are likely to be farther apart due to exchange effects. The coefficient must be chosen empirically until insights into the nature of potential fluctuations have been gained.

The electrostatic interaction energy of electron "$i$" due to other electrons including correlations is given by

$$\Sigma' J_{ij}^{corr} = \int V_i^{eff}(\mathbf{r_1}) |\psi_i(\mathbf{r_1})|^2 d\mathbf{r_1} = \sum_{j \neq i} J_{ij} - \int c_i(\mathbf{r_1}) V_{\Sigma'i}^{\sigma}(\mathbf{r_1}) |\psi_i(\mathbf{r_1})|^2 d\mathbf{r_1} \tag{10}$$

and hence, the correlation energy of electron "$i$" is given by

$$E_i^{corr} = -\int c_i(\mathbf{r_1}) V_{\Sigma'i}^{\sigma}(\mathbf{r_1}) |\psi_i(\mathbf{r_1})|^2 d\mathbf{r_1} \tag{11}$$

The orbital energy is lowered due to electron correlations and is given by

$$\varepsilon_i^{corr} = \varepsilon_i + E_i^{corr} \tag{12}$$

where both $\varepsilon_i$ and $E_i^{corr}$ are negative. The ground state energy is obtained as

$$E_{gs} = \sum_i \varepsilon_i^{corr} - \frac{1}{2} \sum_i \sum_j (J_{ij} - K_{ij}) - \frac{1}{2} \sum_i E_i^{corr} \tag{13}$$

Hence, the total correlation energy, is given by

$$E^{corr} = \frac{1}{2} \sum_i E_i^{corr} \qquad (14)$$

The spread of the orbital energy, $\Delta \varepsilon_i$, is equal to $E_i^{corr}$ determined from Eq. (11) with $c_i(\mathbf{r_1}) = 1$. The higher energy unoccupied states do not play any role in determining electron correlations. In the HF method, the anti-symmetry of wavefunctions (or Pauli exclusion principle) provides some measure of correlation among electrons of same spin resulting in an "exchange hole" surrounding each electron (1,2). In the method described above, electrons of either spin are farther than on average (Eq. (9)) suggesting a "correlation hole" surrounding an electron. This is consistent with the nature of Coulomb interaction, which is independent of spin.

In the HF approximation, the electron – electron interaction term in the Hamiltonian, $1/|\mathbf{r_1} - \mathbf{r_2}|$, is exact but the wavefunction (single Slater Determinant) approximate, due to which it becomes an averaged interaction for the single electron Hamiltonian. This shows that an exact two-particle operator becomes an approximate one-particle operator when the wavefunction is approximate. The above method can be considered to be a correction to this approximate one-particle operator. Within the framework of independent electron approximation, it is equivalent to a first-order perturbation correction to orbital energies.

Of all the methods to determine electron correlation energy, the Configuration Interaction (CI) method (1,2) is conceptually the simplest and can be considered to be a natural extension of the Hartree-Fock method. This is because it is known that the electron-electron interaction term $1/|\mathbf{r_1} - \mathbf{r_2}|$, is exact but the wavefunction (single Slater Determinant) approximate in the HF method. Therefore, the natural course of action would be to expand the true wavefunction in a series of determinants, in which the HF wavefunction would be the first term.

The present method can be considered to be another natural extension of the Hartree-Fock method to incorporate the effects of electron correlations. The natural course of action after determining the average potential is to evaluate its standard deviation. Combined with an empirical constant, this allows an effective potential to be estimated that is closer to the true value than the average (HF) potential. Therefore, this method can also be considered to be a natural extension of the Hartree-Fock method. In this method, the overestimate of the electron-electron repulsion energy is corrected, rather than the wavefunction as in the Configuration Interaction method.

The present method determines electron correlation energy within the single electron approximation and is different from the GW approximation (12-15) and Density Functional approaches (6-11). It does not require any knowledge of Green's functions or pair-correlation functions. It is this conceptual simplicity that can make this method widely accessible and applicable. It is limited in scope as its objective is to determine correlation energy rather than to provide a better theoretical description of electron correlation. Towards this end, it requires the use of an empirical constant. Purely theoretical approaches to electron correlations need to avoid any reliance on empirical constants. To the best knowledge of the author, the method of this paper is not to be found in existing literature.

It is also clear that the method is general and not restricted to electrons. In any system of particles with inter-particle interaction $V(\mathbf{r_1},\mathbf{r_2})$, the potential at any position will fluctuate. The first attempt to solve the Schrodinger's equation usually assumes that the particle moves in an average potential due to other particles, which can be called the mean-field approximation. This paper shows that in addition to the average, the standard deviation of the potential due to other particles can be determined. This allows the spread of single particle energies about their mean-field values to be specified, which provides an indication of the magnitude of correlation energy. In addition, using information about the average and standard deviation along with an empirical constant, an effective potential that is closer to the true value than the average potential can be estimated. The difference in potential energies gives the correlation energy. Conceptually, the way beyond mean-field theory is clear even before knowing the details of the interaction potential $V(\mathbf{r_1},\mathbf{r_2})$. Therefore, this method can be considered to be a universal first step beyond mean-field theory.

Email: prasanna@iitb.ac.in